# Ge virtual substrates for high efficiency III-V solar cells: applications, potential and challenges

Iván García[a], Manuel Hinojosa[a], Iván Lombardero[a], Luis Cifuentes[a], Ignacio Rey-Stolle[a], Carlos Algora[a]
Huy Nguyen[b], Stuart Edwards[b], Aled Morgan[b], and Andrew Johnson[b]

[a] Instituto de Energía Solar, Universidad Politécnica de Madrid, 28040 Madrid, Spain
[b] IQE plc, Pascal Cl, St. Mellons, Cardiff, CF3 0LW, United Kingdom

*Abstract* — Virtual substrates based on thin Ge layers on Si by direct deposition have achieved high quality recently. Their application to high efficiency III-V solar cells is analyzed in this work. Replacing traditional Ge substrates with Ge/Si virtual substrates in standard lattice-matched and upright metamorphic GaInP/Ga(In)As/Ge solar cells is feasible according to our calculations using realistic parameters of state-of-the-art Ge solar cells but with thin bases (< 5μm). The first experimental steps are tackled by implementing Ge single-junction and full GaInP/Ga(In)As/Ge triple-junction solar cells on medium quality Ge/Si virtual substrates with 5μm thick Ge layers. The results show that the photocurrent in the Ge bottom cell is barely enough to achieve current matching with the upper subcells, but the overall performance is poor due to low voltages in the junctions. Moreover, observed cracks in the triple-junction structure point to the need to reduce the thickness of the Ge + III-V structure or using other advanced approaches to mitigate the thermal expansion coefficient mismatch effects, such as using embedded porous silicon. Next experimental work will pursue this objective and use more advanced Ge/Si virtual substrates available with lower threading dislocation densities and different Ge thicknesses.

## I. INTRODUCTION

Investigating alternative substrates for the manufacturing of high efficiency III-V multijunction solar cells is a current topic of intensive research. In addition to the fabrication cost, the availability of the materials used is also a concern. The case of Ge is of particular importance, since it is used for the most mature multijunction solar cells in production [1]. Therefore, obtaining "virtual substrates" based on Ge deposited on Si substrates is an appealing approach. Ge/Si templates conceptually allow the integration of most of the already developed high efficiency III-V multijunction structures, since the lattice constant of Ge is similar to GaAs. The use of SiGe metamorphic buffers to fabricate these templates is one option explored in depth, achieving low threading dislocation densities (TDDs) around $1·10^6$ cm$^{-2}$ and demonstrating promising performances in single and dual-junction GaInP/GaAs solar cells [2], [3]. A disadvantage of this approach is the thick (> 10 μm) metamorphic buffer required. The intermediate chemical-mechanical polishing step used to eliminate the surface cross-hatch roughness, reported by some authors [3], also adds complexity and cost to the fabrication of virtual substrates based on SiGe buffers.

Otherwise, the direct growth of thin Ge films on Si substrates is under development. A two-step deposition process was presented almost two decades ago, which attained a TDD of $2·10^7$cm$^{-2}$ and could be reduced down to $2.3·10^6$ cm$^{-2}$ using selective area growth [4]. More recently, the potential of these virtual substrates for the fabrication of high efficiency III-V solar cell has been experimentally demonstrated by growing GaAs and GaInP solar cells, with promising results [5], [6]. Given the thin Ge layers used (< 5 μm typically), this approach offers an attractive potential for low substrate cost and eases on material scarcity concerns.

In this work we first examine the potential behind using Ge virtual substrates as an alternative to standard Ge and GaAs substrates from a technical perspective. Different possible applications for these virtual substrates are discussed, including standard lattice-matched and upright metamorphic triple-junction solar cells using the Ge virtual substrate as bottom junction, and high efficiency inverted solar cells, where the virtual substrate is meant to be removed and, ultimately, recycled, during the processing of the solar cell. Then, our first experimental results toward the application of these virtual substrates to the fabrication of standard lattice-matched GaInP/Ga(In)As/Ge solar cells are presented and discussed.

## II. METHODS

The modeling and simulation work presented here has been carried out using the commercial TCAD tool Silvaco Atlas. Optical calculations were performed using the transfer matrix method (TMM) included in this software package and using our own developed code. The silicon substrate was only considered for optical purposes. The carrier recombination loss at the Ge/Si interface has been taken into account by including an effective interface minority carrier recombination velocity in the models.

III-V structures were grown in a horizontal, low-pressure, research-scale AIX200/4 reactor. The precursor molecules used where TMGa, TMIn, TMAl, AsH$_3$, PH$_3$, DMZn, CBr$_4$, DTBSi and DETe. A Laytec 2000 in-situ R and RAS monitoring tool was used. The Ge substrates used are 170 μm thick, Ga-doped to around $6 \cdot 10^{17}$ cm$^{-3}$ and with a miscut of 6º towards the nearest (111) plane. Solar cell devices were fabricated using standard photolithography, wet etching techniques, and gold electroplated contacts. External quantum efficiency (EQE) and I-V curves were taken using conventional methods as explained elsewhere [7].

The Ge/Si virtual substrates are fabricated by IQE plc company using silicon substrates with a 6º miscut towards the nearest (111) plane. Germanium epitaxial layers were grown in a single wafer ASM Epsilon epi reactor by reduced pressure chemical vapour deposition (RP-CVD) using H$_2$ as a carrier gas with GeH$_4$ as the source gas and B$_2$H$_6$ for dopant with a growth temperature of less than 900°C. The nominal doping level of the Ge layers is $5 \cdot 10^{17}$ cm$^{-3}$, but they exhibit some non-uniformities in the growth direction, achieving values as high as $2 \cdot 10^{18}$ cm$^{-3}$ at some points. The virtual substrates used for the results presented here correspond to Generation 4, with a thickness of 5 μm and which exhibit a threading dislocation density (TDD) of $4 \cdot 10^6$ cm$^{-2}$. Generations 5 and 6, featuring TDDs below $1 \cdot 10^6$ cm$^{-2}$ are now available and being used for new experiments in this research line.

The structures grown on Ge and Ge/Si substrates in this work are lattice-matched and based on an initial GaInP nucleation layer. The triple-junction solar cell (3JSC) structures used are described in detail elsewhere [8].

The presence of threading dislocations has to be taken into account in any application of the Ge/Si virtual substrates, since they propagate through the epilayers grown on top and affect the minority carrier properties. However, in this work we also pay attention to other aspects that could limit the applicability of these templates, such as the thickness of the Ge layer, the surface readiness for III-V materials nucleation, etc. Other common issue when integrating Si substrates and III-V epi is the thermal expansion coefficient mismatch. The Ge layers are thin enough so that the thermal stress tests carried out on these virtual substrates do not show any issues. Thicker III-V structures grown on top can cause problems in this sense, as shown in the following sections. In addition to lower TDDs, next generation Ge/Si virtual substrates feature an engineered porous Si layer at the interface, which can provide the necessary ductility to accommodate this thermal mismatch. These substrates are now under investigation [9].

### III. POTENTIAL APPLICATIONS

In this section, we discuss the potential applications of Ge/Si virtual substrates from a theoretical perspective and point out the technical challenges associated.

*A. Lattice-matched GaInP/Ga(In)As/Ge/Si Triple-junction Solar Cells.*

The replacement of Ge substrates with Ge/Si virtual substrates in standard lattice-matched GaInP/Ga(In)As/Ge triple-junction solar cells (3JSC) would achieve both a cost reduction and the elimination of the concerns about Ge scarcity in this technology. These solar cells rely on the Ge substrates to form the 3$^{rd}$ junction and are characterized by the low bandgap of Ge, a typical thickness of ~170 μm and high carrier collection efficiencies. With this, the short circuit currents ($J_{sc}$) attained in these Ge subcells are typically above 50% higher than needed to be current matched to the other junctions. The open circuit voltage ($V_{oc}$) in these cells is mostly defined by the emitter properties and the bulk recombination properties in the base [7].

In the virtual Ge/Si substrates contemplated here, Ge is the only photoactive layer, since a negligible fraction of absorbable photons can reach the silicon substrate, given the Ge layer thicknesses considered, conversely to previous modelling work by other authors [10]. Still, the Ge photoactive layer is only a few microns thick. Moreover, although the Ge/Si interface is expected to be type-II with a large conduction band offset [11], which would serve as a good minority carrier barrier, this interface is expected to be highly recombining, due to the misfit dislocations formed there (see Fig. 1). These new characteristics of the substrate that must be used to form the 3$^{rd}$ junction raise questions about the ability to achieve enough photocurrent in this junction or mitigating the effects on the $V_{oc}$ of the highly recombining back interface. In fact, while the emitter properties typically limit the $V_{oc}$ in the Ge subcell of triple-junction solar cells [7], [12], this may not be the case for Ge/Si substrates.

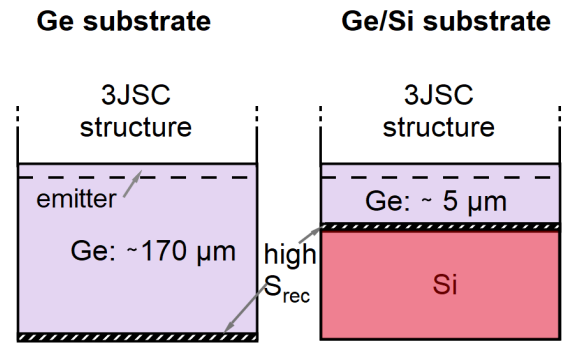

Fig. 1. Sketch comparing the bottom cell structure in a GaInP/Ga(In)As/Ge solar cell grown on Ge and Ge/Si substrates.

In this context, we present here an analysis of the potential of a Ge/Si virtual substrate to function as the bottom cell in a lattice-matched GaInP/Ga(In)As/Ge triple-junction solar cell. We calculate the electrical performance for Ge thicknesses up to 10 μm and for a wide range of back surface recombination velocities. For this, we have done a modelling and simulation study, using the numerical approach explained in the Methods section. The full 3JSC optical stack is included in the model, with a GaInP top cell and Ga(In)As middle cell with thicknesses of 1000 and 4000 nm, respectively. We have used the material properties of our state-of-the-art Ge subcells, including the bulk minority carrier properties (mobilities and lifetimes) in the emitter and base, and the emitter front surface recombination [7], [8], [13]. Assuming the same properties in the case of the Ge/Si virtual substrate, we vary the thickness of the base layer and the back surface recombination properties (represented by the back surface recombination velocity, $S_{rec,Ge/Si}$). In Fig. 2, the resulting contour plots for the $J_{sc}$ and $V_{oc}$ obtained using the AM1.5d-G173 solar spectrum are shown. Before discussing these results, please note that the typical current density in a 3JSC is around 14.5 mA/cm$^2$ at 1-sun for this solar spectrum.

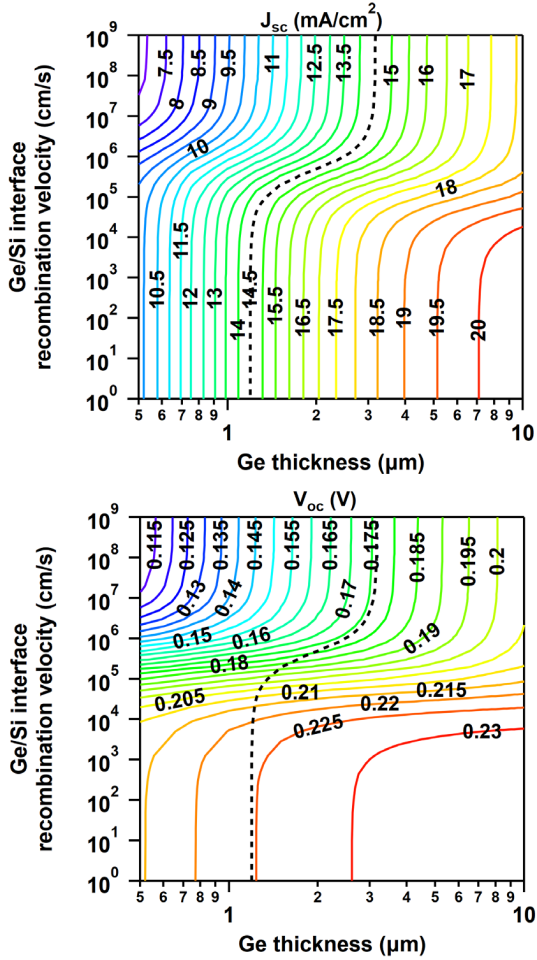

Fig. 2. Calculated contour plots of the $J_{sc}$ and $V_{oc}$ vs the Ge/Si interface recombination velocity and the Ge subcell thickness. The black dashed contours correspond to a $J_{sc}$ of 14.5 mA/cm$^2$. The cells are modelled with anti-reflection coating and the solar spectrum used is the AM1.5d-G173.

The dashed contour in Fig. 2-top corresponds to the 14.5 mA/cm$^2$ $J_{sc}$, and indicates that the minimum thickness of the Ge bottom cell in order not to limit the overall $J_{sc}$ of the 3JSC is between ~1 and ~3 μm, depending on the value of the back surface recombination velocity. This is a convenient range of minimum thicknesses: typical Ge/Si substrates we work with are in the range of 3-5 μm Ge layer thickness and having room for further thinning can be useful to mitigate issues related to thermal expansion coefficient mismatch, as explained in the following sections.

Concerning the $V_{oc}$ contours, shown in Fig. 2-bottom, thinning the Ge in the range shown imply a much more significant $V_{oc}$ loss when the $S_{rec\ Ge/Si}$ is high. For example, thinning the Ge from 5 μm to 1 μm with high $S_{rec\ Ge/Si}$ causes a $V_{oc}$ drop of around 50 mV, but only 10 mV for low $S_{rec\ Ge/Si}$. Therefore, selecting the right Ge layer thickness implies taking into account not only the minimum photocurrent in the bottom cell but also the possible $V_{oc}$ losses associated.

If the $S_{rec\ Ge/Si}$ is high in the Ge/Si substrates (a point that is expected but not corroborated yet), lowering it by using a BSF in the Ge layer (for example using a highly doped region) could widen the range of appropriate thicknesses in the Ge layer, or used to increase the $V_{oc}$. For example, decreasing the $S_{rec\ Ge/Si}$ in a 2.5 μm thick Ge cell would increase the $V_{oc}$ by around 45 mV. It could also enable using a Ge thickness down to ~1 μm at a minimal cost in $V_{oc}$ of around 5 mV.

As a last note, the calculations presented here are based on realistic Ge SJSC properties of as-grown devices, with no further structures grown on top. We know that our MOVPE process for full 3JSC degrades the Ge bottom cell significantly, affecting both the carrier collection efficiency and recombination currents [7], [14]. However, we decided to do the study using the parameters of as-grown Ge SJSC to be more general since other authors do not report carrier collection and $J_{sc}$ losses during the growth of the 3JSC. A similar study, not shown here for brevity, using the parameters of Ge subcells in 3JSC (with worse minority carrier properties in the emitter) results in similarly shaped contour plots but the range of useful Ge thicknesses changes to 2-5 μm, which is still appropriate for the typical Ge/Si virtual substrates we work with.

*B. Other MJSC on Active Ge/Si Virtual Substrates.*

Besides the current-mismatched/lattice-matched approach, other architectures have been developed by several authors to reduce the impact of the excessively low bandgap of Ge on the overall efficiency potential of the MJSC. On the one hand, by using metamorphic structures, the upper subcell bandgaps are lowered to attain a higher current/lower voltage MJSC. The extreme case corresponds to the current-matched MJSC with three and more junctions on Ge [15], while intermediate cases have been also developed, achieving a high maturity and being in production now [16]–[18]. On the other hand, lattice-matched and current-matched 4JSC featuring a subcell made of 1 eV dilute nitride material is also under development [14]. All these cases have in common a narrowing of the spectral absorption band of the Ge subcell, which reduces the bottom cell photocurrent decrease affordable when replacing the Ge substrate with Ge/Si virtual substrates.

Obviously, in the case of current-matched structures, the photocurrent drop in the Ge bottom cell by using a Ge/Si substrate will inevitable entail a performance loss. However, assessing the tradeoff between this loss and the economic benefits of Ge/Si substrates is beyond the scope of this work. For the intermediate cases where the Ge bottom cell is still producing excess photocurrent, it is interesting to quantify the room for thinning of the Ge bottom cell without limiting the MJSC $J_{sc}$. Fig 3-bottom shows the calculation of the minimum Ge thickness needed in a 3JSC to achieve current matching for a range of upper subcells photocurrents attained by varying their bandgaps. The top graph shows the three EQE corresponding to the cases of lattice-matching, commercial upright metamorphic 3JSC and current matched with a standard Ge substrate. The EQE of the Ge bottom cells shown corresponds to the thinned cases that achieve current matching (see the thicknesses used in the labels). The Ge solar cell parameters used are constant for all the thicknesses and correspond to the state-of-the-art Ge bottom cells shown in the previous section and using a negligible Ge/Si interface recombination velocity. This is therefore an optimistic assessment, since the bottom cell is expected to have a lower quality for thin Ge/Si virtual substrates. The minimum thicknesses for current matching shown in Fig 3-bottom are then a lower limit. This graph shows that the thicknesses required for lattice-matched and commercial structures are in the range of 1 to 3 μm, indicating that the use of Ge/Si virtual substrates is conceptually feasible in these cases.

As the top cell bandgap decreases, it removes photons from the middle cell which is forced to decrease its bandgap increasingly faster as it transitions through the water absorption band in the AM1.5d-G173 spectrum and then through a region with low photon content. This causes that the rate of minimum thickness increase rises as the 3JSC $J_{sc}$ increases, as shown in Fig 3-bottom. This limits the applicability of the Ge/Si substrates for upright metamorphic solar cells with low bandgaps. Significantly different results are expected when using other spectra such as AM0 for space applications.

*C. High Efficiency Upright and Inverted Solar Cells Using Inactive Ge/Si Virtual Substrates.*

Highest efficiency III-V multijunction solar cells, including inverted metamorphic, wafer bonded and dilute-nitride based approaches, use GaAs substrates. Despite being cheaper, using Ge substrates instead is not common, probably because the cost advantage is counterbalanced by issues such as the added complexity of the III-V on Ge nucleation routines needed and associated reactor conditioning [19]. However, with the cost reduction of the substrate attained with the Ge/Si templates, the replacement of GaAs substrates by Ge/Si virtual substrates may become worthwhile.

Some issues must be taken in consideration when using Ge/Si templates for this application. First, the removal of the substrate by selective chemical etching is not as straightforward as when using GaAs substrates. Controllable, fast and selective etching processes are complicated to achieve. Our efforts so far to demonstrate an efficient selective removal of Ge/Si virtual substrates have not been fruitful. On the other hand, substrate reuse by ELO is expected to be suitable for Ge/Si substrates, and experiments in this direction are planned for the near future.

Ge autodoping and solid phase diffusion must be considered too. The lessons learned on upright structures must be rethought given the different active parts of the structure affected more strongly in inverted structures (highly sensitive GaInP [20] is grown first), the higher thermal loads in structures with metamorphic buffers, high temperature layers (AlGaInP), added junctions, etc. Ge autodoping is caused by the Ge etched from the back of the substrate, which is why this effect is typically minimized by using $SiN_x$ coated substrates. Virtual Ge/Si substrates have no Ge surface exposed, as the Ge layer is covered by the Si substrate. We have grown GaInP structures on Ge/Si to test the effect of Ge outdiffusion and have found no significant presence of Ge in the GaInP layer by SIMS. Concerning the Ge solid phase diffusion, previous studies show that its effect on 3JSC can be removed by using the proper Ga(In)As overbuffer thickness [21]. We are now investigating the case of inverted metamorphic structures, where the structure is subjected to heavier thermal loads.

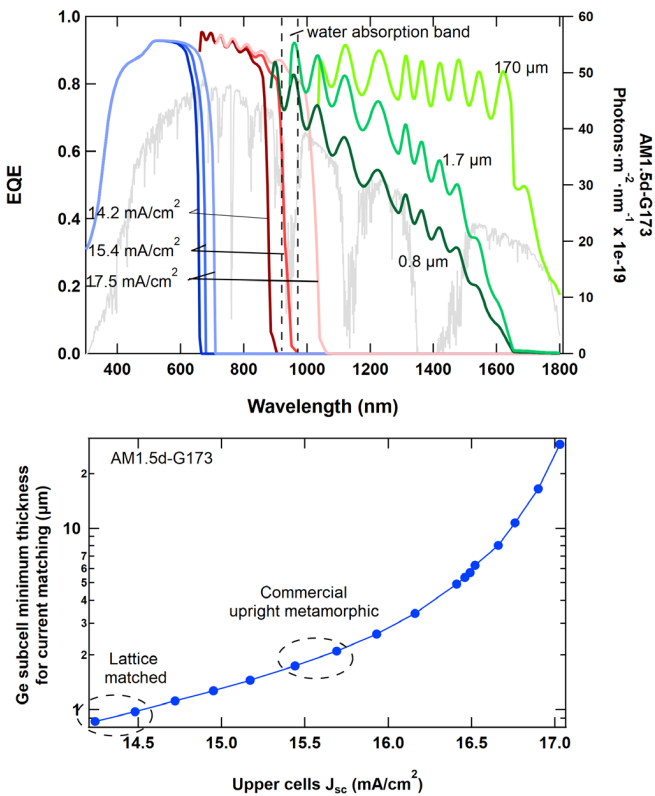

Fig. 3. Top: EQE calculated for GaInP/Ga(In)As/Ge 3JSC for different bandgaps of the upper junctions that produce current matching. The labels indicate the $J_{sc}$ obtained in each case and the thickness of the Ge bottom cell. Bottom: minimum Ge bottom cell thickness required for current matching as the $J_{sc}$ of the 3JSC increases by decreasing the bandgaps of the upper junctions.

## IV. EXPERIMENTAL TRIPLE JUNCTION SOLAR CELLS

In this section we present the first experimental steps towards the development of standard lattice-matched GaInP/Ga(In)As/Ge on Ge/Si virtual substrates, whose feasibility was analyzed in the previous section. The results presented here were obtained using 5 μm thick Ge layers in the Ge/Si substrates, but the study of thinner substrates is underway.

### A. Nucleation of III-V Materials on Ge/Si Virtual Substrates.

The implementation of 3JSC on Ge/Si substrates requires, first, that they allow a proper nucleation of III-V materials for the formation of the GaInP window layer of the Ge bottom cell and for the growth of the upper subcells. High quality nucleation and growth of III-V materials on Ge by MOVPE has been studied in depth and is well known (a complete summary can be found in [19] and references therein). The observation of the in-situ reflectance anisotropy spectroscopy (RAS) signal during the growth of GaInP nucleation layers can be used to determine the readiness of the surface and the quality of the nucleated layers, as shown in [22].

In Figure 4, the transient RAS and R signals taken at 2.1 eV during the GaInP nucleation and Ga(In)As overbuffer steps is shown for the case of using a Ge/Si virtual substrate. The RAS signature observed at the first nanometers of GaInP growth is indicative of the quality of the wafer surface regarding nucleation of GaInP [19]: the lower the amplitude of this signature, the higher the quality. As can be observed, the amplitude measured is very small, indicating a pristine Ge surface producing an excellent nucleation. The signal recorded during the subsequent growth of the GaInP nucleation and Ga(In)As overbuffer shows a specular surface with no apparent roughening.

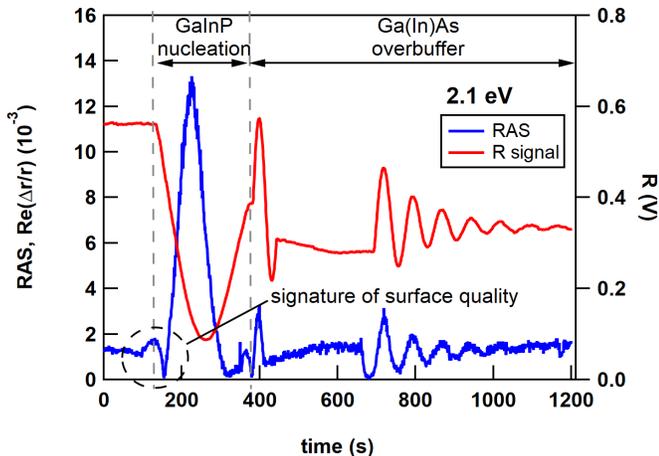

Fig. 4. In-situ characterization of the growth surface during nucleation of GaInP on Ge/Si virtual substrates and subsequent growth of Ga(In)As overbuffer layer. The R signal is the raw voltage provided by the photodetector and is therefore proportional to the real reflectivity of the sample.

### B. Ge Solar Cells Formed on Ge/Si Virtual Substrates.

Using the optimized nucleation routine explained, complete solar cell structures were grown. First, the Ge bottom cell is tackled. Ge single-junction solar cells (SJSC) were fabricated using Ge/Si virtual substrates with 5 μm thick Ge layers, and using standard Ge substrates (see Methods section for details). With these structures, solar cells were fabricated, characterized and compared. The EQE results are shown in Fig 5.

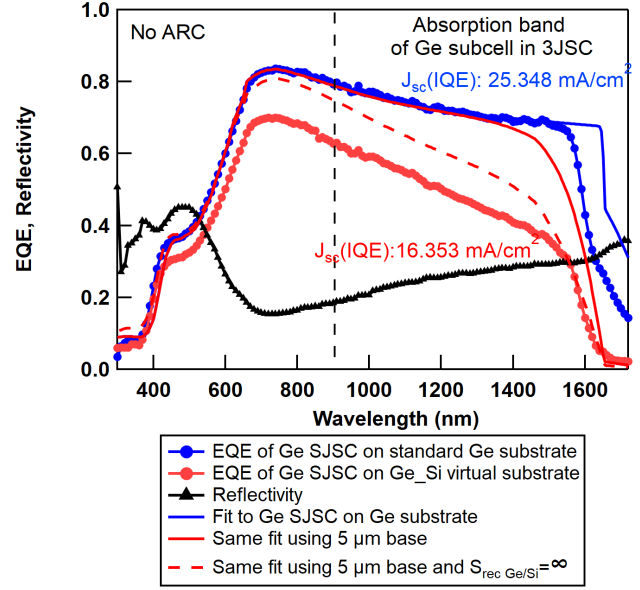

Fig. 5. Symbols: experimental EQE and reflectivity of Ge solar cells fabricated on Ge and Ge/Si substrates; lines: fit to the EQE of the standard Ge substrate (170 μm) and same fits but changing first the base thickness to 5 μm, and then also using an infinite $S_{rec\ Ge/Si}$.

A significantly lower EQE is obtained for the virtual substrate case. In order to clarify the cause, the EQE of the Ge subcell on standard substrate was fitted and then recalculated using the same fitted parameters but a 5 μm base layer. The results, plotted in Fig 5, show first an almost perfect fit to the experimental EQE for the Ge SJSC fabricated on a standard Ge substrate (the discrepancy at long wavelengths is caused by free carrier absorption effects not included in the model [23]). However, it also reveals that just lowering the Ge thickness to 5 μm with the same parameters should produce changes in the EQE only at wavelengths above 1400 nm, but the experimental EQE of the Ge SJSC fabricated on the Ge/Si virtual substrates shows a considerable drop at all wavelengths. This indicates an inferior minority carrier collection due to worse minority carrier properties in the Ge material and/or a high back surface recombination velocity. It seems unlikely that a TDD of $4 \cdot 10^6$ cm$^{-2}$ could have such a strong effect on the EQE, at least by comparison with metamorphic GaInAs solar cells with similar bandgap and TDD [24]. On the other hand, a high back surface recombination velocity does not justify the EQE results, as shown in Fig. 5: the dashed red line EQE is obtained using the parameters of the 5 μm case and infinite $S_{rec\,Ge/Si}$. Therefore, the quality of the Ge material deposited seems to be significantly worse than bulk Ge in conventional substrates.

Concerning the operation of these Ge bottom cells in complete 3JSC, it is essential that they achieve the required photocurrent in order not to limit the $J_{sc}$ in the multijunction solar cell. In Fig 5, the $J_{sc}$ calculated using the IQE for the absorption band of the Ge subcells and the AM1.5d-G173 spectrum is detailed with the labels. The value obtained for the Ge subcell on Ge/Si virtual substrates is barely enough to achieve current matching with the other subcells in a 3JSC: if we apply a correction factor to take into account the typical reflection losses in the bottom cell of these 3JSC (around 10-15%), the $J_{sc}$ obtained is approximately 14.5 mA/cm$^2$, as needed for current matching with the upper subcells.

Concerning the light I-V curves measured (Fig. 6), the Ge subcells fabricated on standard substrates exhibit state-of-the-art performance, with an $V_{oc}$ slightly over 0.25 V. As for the Ge/Si virtual substrates, the $V_{oc}$ obtained is around 100 mV lower, in line with the worse material quality deduced from the EQE data. Moreover, these cells exhibit a leaky behavior, with no flat region around the 0V bias point and a current increasing steadily at negative voltages. While the detailed origin of this behavior is under study now, we hypothesize it could be a combination of a low reverse breakdown voltage and shunts originated by TD or other defects.

All in all, these results show that, despite the photocurrent obtained can be enough to achieve current matching, the material quality needs to be improved to reduce the voltage losses in Ge solar cells made on Ge/Si virtual substrates.

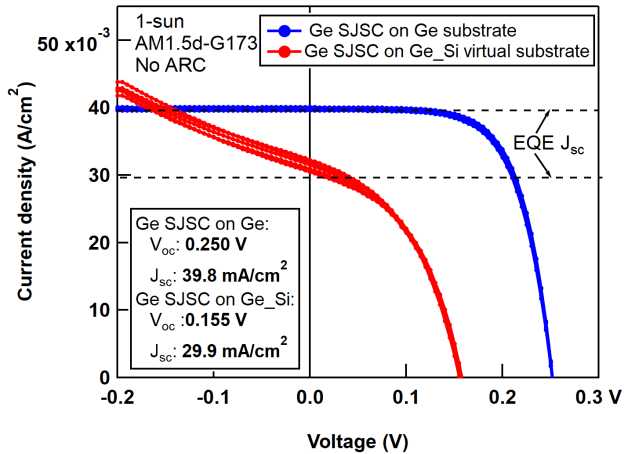

Fig. 6. Light I-V curves of the Ge solar cells analyzed, taken at 1-sun under the AM1.5d-G173 spectrum.

*C. Triple-junction Solar Cell on Ge/Si Substrates.*

Complete triple-junction solar cell structures were also grown on Ge and Ge/Si substrates. One of the first issues encountered in these first steps towards the use of Ge/Si virtual substrates for 3JSC is the formation of cracks along the epilayer surface in these structures. In Figure 7 an example of these cracks in a 3JSC structure is shown. The cracks are associated with the growth of a thick III-V structure, which is added to the Ge layer in the virtual substrate, and the thermal expansion coefficient mismatch between the Si substrate and the Ge and III-V layers. The samples used for this study were grown on Ge/Si substrates with a Ge thickness of 5 μm. These substrates showed a few cracks at the edges before the growths. However, annealing experiments carried out on them, using a temperature of 700ºC for 10 min, did not induce the formation of any additional crack. Conversely, after the growth of the 3JSC structures, with a total thickness of the III-V structure of around 5 μm, new cracks appeared easily when manipulating the samples (for example when dicing them). This could be expected, as the III-V layer thickness grown on Si that produces cracks was estimated to be 3 μm by other authors [25]. In fact, this problem was not observed in the Ge bottom cells grown on the same Ge/Si virtual substrates. The use of Ge/Si virtual substrates with thinner Ge layers and minimizing the total thickness of the III-V structure is now being investigated to reduce the formation of cracks. However, using thin Ge layers in the Ge/Si substrates will require improving the photovoltaic quality of these layers, as shown in Section II. Still, as commented above, the most promising approach being investigated is the use of Ge/Si virtual substrates featuring porous Si layers in order to provide the required ductility at the Ge/Si interface to accommodate the differences in thermal expansion coefficients.

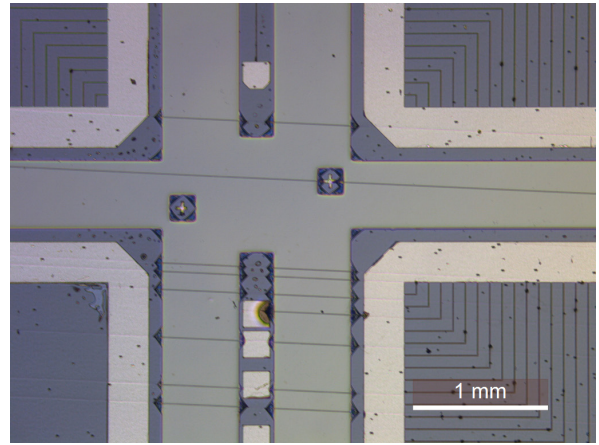

Fig. 7. Front view of the 3JSC showing cracks running along the surface of the epilayers (darker areas) and substrate (lighter areas).

3JSC solar cell devices were fabricated using the structures grown, and their EQE and light I-V curves measured. Fig. 8 shows the EQE for the devices without ARC. Several aspects can be pointed out. The Ge bottom cell could not be measured for the 3JSC grown on Ge/Si substrates, given the leaky behavior of this junction [26]. The Ga(In)As middle cell of this 3JSC shows a significantly lower EQE than for the standard Ge substrate case. Interestingly, the GaInP top cell shows a virtually identical EQE in both cases.

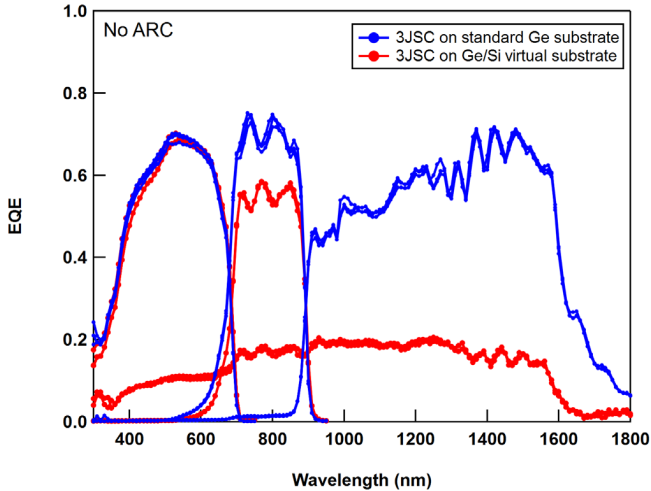
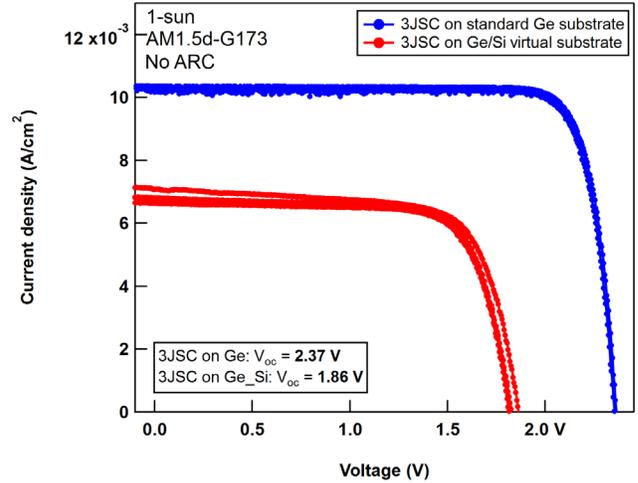

Fig. 8.  EQE of the 3JSC fabricated on Ge and Ge/Si substrates. The measurement of the Ge bottom cell in the latter case was not possible due to its leaky behavior.

Fig. 9.  Light I-V curves of the 3JSC solar cells analyzed, taken at 1-sun under the AM1.5d-G173 spectrum.

Fig. 9 shows the light I-V curves obtained. As expected, the performance of the 3JSC grown on Ge/Si substrates is poor. On top of the lower $J_{sc}$ caused by the low EQE response of the middle and, possibly, bottom cells, we observe a dramatic $V_{oc}$ loss of around 500 mV with respect to the standard Ge substrate case. Part of this $V_{oc}$ loss is attributable to the Ge bottom cell, as shown in the previous section. Note that the loss observed in Ge SJSCs could be amplified due to the effect of the growth of the 3JSC structure [7], [14]. Still, to explain the 500 mV loss in the 3JSC, the $V_{oc}$ must be degraded also in some or all the other two junctions.

The expected TDD of the Ge/Si templates used, around $4·10^6$ cm$^{-2}$, could be partly responsible of these results, but the comparison with inverted metamorphic 3JSC structures with similar TDD developed in our laboratory, and with GaAs and GaInP SJSCs grown on Ge/Si virtual substrates by other authors [5], [6], suggest that there must be additional causes. These could be linked to the mechanical instability of the semiconductor structure to temperature variations. We hypothesize with the possibility that the temperature ramps used for the growth of the 2$^{nd}$ tunnel junction in the structure, right after the thick (~ 4μm) Ga(In)As middle cell is deposited, could result in mechanical defects (for example internal cracks) affecting this subcell. In fact, other authors have reported much better performances in thin (~ 1 μm) GaAs SJSC grown on Ge/Si substrates with similar TDD [5]. Concerning the GaInP top cell, it is possible that, being thinner and subjected only to a temperature ramp-down during the final cooling step after the growth process, it is less exposed to the formation of internal cracks of this kind. However, regardless of the actual explanation of these experimental results, it becomes apparent that the Ge/Si substrates need to be engineered in order to improve their suitability to serve as growth template and bottom subcell in GaInP/Ga(In)As/Ge 3JSCs.

V. CONCLUSIONS.

The potential of Ge/Si virtual substrates to be used in standard lattice-matched and upright metamorphic GaInP/Ga(In)As/Ge triple-junction solar cells has been theoretically demonstrated, assuming a realistic Ge bottom cell photovoltaic quality similar to the standard Ge substrates case, and a range of back surface recombination velocities. The experimental results carried out using medium quality Ge/Si virtual substrates demonstrate a good nucleation of III-V materials on these substrates but show that their photovoltaic quality needs to be improved to match the performance obtained in Ge single-junction and complete triple-junction solar cells using standard Ge substrates. Better Ge layer qualities could also provide some room for their thinning in order to minimize the observed deleterious effects of the thermal expansion coefficient mismatch with the silicon substrate. For the same purpose, using embedded porous silicon layers is being investigated. A similar experimental approach using higher quality Ge/Si templates, exhibiting threading dislocation densities below $5·10^5$ cm$^{-2}$, together with the fabrication of inverted structures with substrate reuse by ELO, is planned for the next steps towards developing high efficiency multijunction solar cells on Ge/Si virtual substrates.

ACKNOWLEDGMENTS

This project has been funded by the Spanish Ministerio de Ciencia, Innovación y Universidades with the projects TEC2015-66722-R and RTI2018-094291-B-I00, and by Comunidad de Madrid with project MADRID-PV2 (S2018/EMT-4308). M. Hinojosa is funded by the Spanish MECD through a FPU grant (FPU-15/03436) and I. García is funded by the Spanish Programa Estatal de Promoción del Talento y su Empleabilidad through a Ramón y Cajal grant (RYC-2014-15621).